# Transition from acoustic plasmon to electronic sound in graphene

**Authors:** David Barcons Ruiz[1], Niels C.H. Hesp[1], Hanan Herzig Sheinfux[1], Carlos Ramos Marimón[1], Curdin Martin Maissen[2], Alessandro Principi[3], Reza Asgari[4,5], Takashi Taniguchi[6], Kenji Watanabe[7], Marco Polini[1,8,9], Rainer Hillenbrand[2,10,11], Iacopo Torre[1]*, Frank H.L. Koppens[1,12]*

**Affiliations:**

[1]ICFO-Institut de Ciències Fotòniques, The Barcelona Institute of Science and Technology, Av. Carl Friedrich Gauss 3, 08860 Castelldefels (Barcelona), Spain.

[2]CIC nanoGUNE; 20018 Donostia-San Sebastián, Spain.

[3]Department of Physics and Astronomy, The University of Manchester, M13 9PL Manchester, United Kingdom

[4]School of Physics, Institute for Research in Fundamental Sciences, IPM, Tehran, 19395-5531, Iran

[5]School of Physics, University of New South Wales, Kensington, NSW 2052, Australia

[6]International Center for Materials Nanoarchitectonics, National Institute for Materials Science, 1-1 Namiki, Tsukuba 305-0044, Japan

[7]Research Center for Functional Materials, National Institute for Materials Science, 1-1 Namiki, Tsukuba 305-0044, Japan

[8]Dipartimento di Fisica, Università di Pisa, Largo Bruno Pontecorvo 3, I-56127 Pisa, Italy

[9]Istituto Italiano di Tecnologia, Graphene Labs, Via Morego 30, I-16163 Genova, Italy

[10]CIC nanoGUNE BRTA and Department of Electricity and Electronics, UPV/EHU, 20018 Donostia-San Sebastián, Spain.

[11]IKERBASQUE, Basque Foundation for Science, Bilbao, Spain

[12]ICREA-Institucio Catalana de Recerca i Estudis Avancats, 08010 Barcelona, Spain

*Corresponding authors. Emails: frank.koppens@icfo.eu, iacopo.torre@icfo.eu

**Abstract:**

Fermi liquids respond differently to perturbations depending on whether their frequency is larger (collisionless regime) or smaller (hydrodynamic regime) than the inter-particle collision rate. This results in a different phase velocity between the collisionless zero sound and hydrodynamic first sound. We performed terahertz photocurrent nanoscopy measurements on graphene devices, with a metallic gate in close proximity to the sample, to probe the dispersion of propagating acoustic plasmons, the counterpart of sound modes in electronic Fermi liquids. We report the observation of a change in the plasmon phase velocity when the excitation frequency approaches the electron-electron collision rate. This first observation of the first sound mode in an electronic Fermi liquid is of fundamental interest and can enable novel terahertz emitter and detection implementations.

**Main Text:**

The Fermi liquid paradigm[1,2] is one of the cornerstones of modern condensed matter theory, providing an effective description of the many-body systems whose elementary excitations are weakly-interacting fermionic quasi-particles. Most importantly, the theory of Fermi liquids provides an understanding of why conduction electrons in metals behave essentially as non-interacting particles. Fermi liquids can support collective modes in the form of longitudinal density oscillations that are analogous to sound in classical fluids. The propagation of collective modes in Fermi liquids depends on whether the angular frequency $\omega$ of the mode is higher or lower than the inter-particle collision rate[3] $\tau_{\text{coll}}^{-1}$. Liquid $^3$He, a neutral Fermi liquid, was the first system in which the transition (a change in the velocity and attenuation of the propagating mode) from the first sound ($\omega \ll \tau_{\text{coll}}^{-1}$, i.e. in the hydrodynamic regime) to the zero sound mode ($\omega \gg \tau_{\text{coll}}^{-1}$, i.e. in the collisionless regime) was observed[4].

In electronic Fermi liquids with long-range Coulomb interactions, first and zero sound collapse into a plasmon mode[5]. In such a mode, the smooth crossover from the collisionless to the hydrodynamic regime manifests in the dispersion relation $\omega(q)$ only at subleading order in the wave-vector $q$ of the mode[5] and it is therefore very challenging to observe. However, two-dimensional (2D) electron liquids allow for sufficient screening of the long-range part of the Coulomb interaction by a nearby metallic gate electrode, the first and zero sound reappear[5–7] and a transition between the two can be observed. In 2D electron liquids with screened electron-electron (ee) interactions, the zero sound mode is known as acoustic plasmon and has been extensively studied experimentally in hexagonal boron nitride (hBN) encapsulated graphene devices[8,9]. However, to the best of our knowledge the electronic first sound mode, the closest electronic analogue of ordinary sound, has never been observed experimentally.

In this work, we report the first evidence of a transition between an acoustic plasmon and the electronic first sound. We probe this transition at room temperature (RT) employing a terahertz (THz) source whose angular frequency $\omega$ can be tuned around the ee collision rate $\tau_{\text{ee}}^{-1}$, with $\tau_{\text{ee}}$ being $0.1 - 0.2$ ps in doped graphene[10–12]. Figure 1**A** shows the evolution of the distribution function of electrons during the propagation of a plasmon mode. While for $\omega \gg \tau_{\text{ee}}^{-1}$ the distribution function differs greatly from the equilibrium one, for $\omega \ll \tau_{\text{ee}}^{-1}$ ee collisions have time to smooth the distribution to a circle, leading to a quasi-equilibrium, fluid-like response. This results into a change in the phase velocity of the mode between the two regimes (see Fig. 1**B**).

The plasmon velocity difference has, at long wavelength, an intuitive physical interpretation thanks to an analogy with viscoelastic materials[13]. Materials respond as solids (with an elastic shear force) to shear deformations that are faster than a certain equilibration time scale $\tau_{\text{coll}}$ and as fluids (with a dissipative shear force) to shear deformations slower than $\tau_{\text{coll}}$. This time scale greatly varies (even by orders of magnitude) depending on the material, and diverges for ordered solids. Electrons are no exception to this behaviour with the role of $\tau_{\text{coll}}$ played by $\tau_{\text{ee}}$. In the collisionless regime the elastic shear force, that is not present in the hydrodynamic regime, adds to the Coulomb and pressure forces[13] in sustaining the plasmon mode. This makes the plasmon mode stiffer, increasing its velocity. In order to observe this effect, it is however necessary to screen the otherwise dominant Coulomb force.

The electronic first sound pertains to the hydrodynamic regime, which is characterized by $\tau_{\text{ee}}$ being the shortest timescale of the system[14–16]. Quantitatively, this happens when $\tau_{\text{ee}} \ll \tau, \omega^{-1}, (qv_{\text{F}})^{-1}$, where $\tau$ is the momentum relaxation time, $\omega^{-1}$ is the time over which the phase of the mode changes significantly, and $(qv_{\text{F}})^{-1}$ ($q$ being the wavevector of the mode) is the time it takes to an electron travelling at the Fermi velocity $v_{\text{F}}$ to cross a significant fraction of a spatial oscillation of the mode (the corresponding ranges of frequencies relevant for our experiment are depicted in Fig. 1C). Since both $\tau_{ee}^{-1}$ and $\tau^{-1}$ increase with temperature, typically at different rates, the hydrodynamic regime can only be realized in high-mobility electronic systems for a limited window of experimental conditions[10,17,18]. The hydrodynamic regime has been demonstrated experimentally in encapsulated graphene samples[10,11,19–22] or GaAs/AlGaAs quantum wells[18,23]. The transition between zero and first-sound in 2D electronic liquid was studied theoretically[5–7] using simplified models based on the semi-classical Boltzmann transport equation that fully captures nonlocal effects[9,24]. The magnitude of the difference between the plasmon velocity in the two regimes is controlled by the screening parameter[7][†]

$$\Lambda = \frac{C}{e^2 N} \approx \frac{1}{t_{\text{hBN}}[\text{nm}]\sqrt{|n[10^{12} \cdot \text{cm}^{-2}]|}}, \quad (1)$$

where $C$ is the capacitance per unit area between the electron liquid and the metallic gate, $e$ is the unit charge and $N$ is the electronic density of states at the chemical potential. The second relation holds for the specific case of single-layer graphene with carrier density $n$ and separated from a nearby metallic gate by an hBN spacer of thickness $t_{\text{hBN}}$. Figure 1D shows the values of $\Lambda$ that can be reached as a function of the experimental parameters.

The sought effect is negligible for $\Lambda \approx 0$ but becomes strong for $\Lambda \approx 1$. When the latter condition is reached, the hydrodynamic plasmon velocity becomes even smaller than $v_F$. In the extreme case of very large screening ($\Lambda \gg 1$) the collisionless plasmon velocity tends to $v_F$, while the hydrodynamic plasmon velocity tends to the 2D energy-waves (second sound) velocity[25] $v_F/\sqrt{2}$. The convergence of these two modes to the same limiting velocity can be understood since they both approach charge-neutral oscillations. In the case of the second sound, this happens because of the charge compensation between electrons and holes, while in the case of acoustic plasmons the same happens because of the compensation due to induced image charges in the metallic gate. Based on the theoretical model presented in Ref.[7] and making an approximation that is well-justified in single-layer graphene[26] (i.e. neglecting the first-order Landau parameter $F_1^s$) it is possible to derive (Supplementary Note 4) a simple relation between the collisionless plasmon velocity $v_c$ and the hydrodynamic plasmon velocity $v_h$:

$$v_h = \sqrt{\frac{v_c^2 + v_c\sqrt{v_c^2 - v_F^2}}{2}}, \quad (2)$$

with $v_F$ the (renormalized) Fermi velocity. From this formula we immediately see that the difference between the two velocities is negligible if $v_c \gg v_F$ (corresponding to small values of $\Lambda$) and becomes most important when $v_c \approx v_F$. Even in this extreme case, the relative difference between the two velocities cannot exceed $|v_h - v_c|/v_c \lesssim 29\%$.

**Experiment**

In this work, we probe the transition between the collisionless and the hydrodynamic regime of electrons in graphene by measuring the phase velocity of acoustic plasmons at different angular frequencies close to the expected value of $\tau_{ee}^{-1}$ that we calculated for the specific structures of our experiment (See Fig. 1C and Supplementary Note 7). To this aim, we fabricated two hBN-encapsulated single-layer graphene devices, dubbed device 1 and device 2, respectively, with different gate-graphene separation. The transition effect is observed in device 1, while device 2 is a control device in which this effect is predicted to be negligible. The two devices share the same split-gate configuration depicted in Fig. 1**E**. They consist on hBN/graphene/hBN heterostructures on top of metallic palladium gates. Each metallic gate is split into two halves whose voltages can be controlled separately. This allows the creation of a sharp p-n junction in the sample, which enables the thermoelectric detection of the plasmonic field[27].

---

[†] The definition of $\Lambda$ used here differs slightly from the one used in Ref. [7] in that the density of states $N$ appearing in (1) is the observed or renormalized one, not the bare one. In this work we will denote with a subscript 0 ($N_0$) the bare, non-interacting, quantities.

The only relevant difference between the two devices is the thickness of the bottom hBN spacer ($t_{hBN}$ in Fig. 1**E**) that is chosen to be as small as possible ($t_{hBN}$ = 2.0 nm, leading to a design value of $\Lambda \approx 0.5$ at the carrier densities used in the experiment) in device 1, and larger ($t_{hBN}$ = 11.8 nm, corresponding to $\Lambda \approx 0.08$) in device 2. This means that the Coulomb interaction should be strongly screened in device 1 (quantified by higher values of $\Lambda$ as seen in Fig. 1**D**), where acoustic plasmons are expected to propagate with low velocity (with $v_p/v_F$ reaching values as low as 1.5). This yields a sizable change in $v_p$ between the two regimes. On the contrary, in device 2, the Coulomb interaction is less screened (see Fig. 1**D**) and $v_p/v_F$ never goes below 2.5. This means that, in device 2, $v_p$ is almost the same in the two regimes and no significant transition effect is expected.

We performed THz photocurrent nanoscopy[8] measurements at RT ($T = 295$ K) in a commercial scanning nearfield optical microscope (SNOM). We used a methanol gas laser to measure at four different frequencies ($f = 1.84, 2.52, 3.11, 4.25$ THz). For each laser frequency, we scanned the tip repeatedly along the white dashed lines indicated in Fig. 2**A,B**, for a set of gate voltages V$_1$, while the other gate is kept at a voltage V$_2$ chosen to maximize the photocurrent signal (see Methods). The dominating plasmon launching mechanism differs between the two devices due to the very different vertical confinement of the plasmon (see Ref. [9] and Supplementary Note 1). In device 1, the sharp metallic edge at the junction launches a plane wave propagating perpendicular to it. By scanning the tip perpendicular to the junction, we measure λ-fringes (i.e. fringes with periodicity equal to the plasmon wavelength). In device 2, the tip launches a circular plasmonic wave. By scanning the tip parallel to the junction, we detect λ/2-fringes (i.e. fringes with periodicity equal to half the plasmon wavelength) due to the standing wave originating from the plasmons reflected at the graphene edge and travelling back to the tip[28]. Inverting the measurement direction between both devices results in barely visible fringes. In both cases, the SNOM tip serves as a local probe to rectify the plasmonic field, generating heat, which is then converted into photocurrent at the p-n junction via thermoelectric effect[27].

Figure 2**C,D** display the real part of the photocurrent signal (recorded at the first harmonic of the tip frequency) acquired for both devices at the four studied frequencies, at a carrier density $n \approx 10^{12}$ cm$^{-2}$. Both devices, and in particular device 1, display very high electronic quality with plasmon lifetimes of about $0.5 - 1$ ps (Supplementary Note 5). This is evident from the quality of the data in Fig. 2**C,D,** which show up to 7 clearly visible oscillation fringes. The presence of a good number of fringes is pivotal for a reliable extraction of the plasmon wavelength. Due to the different scanning directions, in device 1 there is an additional decay of the signal along the scanning direction as the tip moves away from the junction[27]. Conversely, in device 2 the tip-junction distance is kept constant but there is a geometric decay due to the cylindrical plasmonic wave radiating away from the tip (Supplementary Notes 1 and 2). The high signal-to-noise ratio typical of our technique and the high mobility of our devices allow to accurately extract the plasmon wavelength $\lambda_p = 2\pi/q_p$ for carrier densities above $0.5 \cdot 10^{12}$ cm$^{-2}$ for device 1, and above $0.3 \cdot 10^{12}$ cm$^{-2}$ for device 2 (see Supplementary Note 2 for the full data sets and the fitting-procedure details).

From these measured $\lambda_p$, we extract the plasmon phase velocity as a function of the gate voltage (Fig. 3**A,B)**, measured with respect to the charge neutrality point, determined by two-probe transconductance measurements, $V_G = V_1 - V_{1,\text{CNP}}$, for device 1 and 2 respectively. We were able to extract the plasmon wavelength with a sufficient degree of accuracy at the four laser frequencies. Measuring at lower laser frequencies was not possible because $\lambda_p$ becomes too large compared with device dimensions, propagation and cooling length, making the extraction of the wavelength not reliable enough. From the measured plasmon velocity we find (see Supplementary Note 3) a smaller capacitance than the one expected from the thickness of the exfoliated hBN flakes in both devices, yielding typical values of $\Lambda$ of $\approx 0.2$ and $\approx 0.04$ in device 1 and 2 respectively. We attribute this discrepancy to air gaps between the metallic gate and the hBN flake (see Supplementary Note 3). This effect can be described as an effective hBN thickness larger than the nominal one. We used this more realistic quantity to locate our devices in the parameter space plotted in Fig. 1**D**.

**Discussion**

The most notable observation is that the two devices show a different dependence of the plasmon velocity with frequency. In contrast to device 2 (Fig. 3**B**), the data of device 1 (Fig. 3**A**) show a clear frequency dependence, with the plasmon velocity slowing down by $\approx 5\%$ from the highest frequency to the lowest. The effect is emphasized in the inset that shows the relative variation with respect to the highest frequency. We will show that our findings are compatible with the transition from the collisionless to the hydrodynamic regime. To this aim, we compare Eq. (2) with our data, by approximating the collisionless velocity $v_c$ with our velocity data at the highest available frequency 4.25 THz and the Fermi velocity with its value calculated at the carrier density of $n = 10^{12}\text{cm}^{-2}$, $v_F \simeq 1.1 \cdot 10^6$ m/s (Supplementary Note 3). For more clarity, instead of applying Eq. (2) directly, we fit the experimental data with a simple one-parameter model derived from the collisionless expression (Supplementary Note 4) $v_c(V_G) = v_F \left(\alpha\sqrt{|V_G|} + 1\right)\left(2\alpha\sqrt{|V_G|} + 1\right)^{-1/2}$ (green solid line in Fig. 3**A**). We then apply Eq. (2) to the fitted curve to obtain our theoretical estimate of the hydrodynamic velocity (green dashed line in Fig. 3**A**). Remarkably, the theoretical line matches well with the data extracted at the lowest available frequency of 1.84 THz. As expected, repeating the same procedure with the control device we find only a very small velocity shift that is compatible with experimental errors, as shown in Fig. 3**B**. In Fig. 3**C,** we display the plasmon velocity for a fixed carrier density $n \approx 10^{12}$ cm$^{-2}$ as a function of the laser frequency. To reduce the uncertainty associated with the use of a single experimental point in voltage, the data points and their error bars have been obtained from fitting the experimental plasmon dispersions in Fig. 3**A,B** to a functional expression which allows to fit the dispersion in both the hydrodynamic and collisionless regime: $v_p(V_G) = \alpha|V_G|^{1/4}(1 + \beta|V_G|)$.

We note that refs. [5–7] also predict a change in the plasmon damping rate $\Gamma \approx \omega\text{Re}\{q_p\}/\text{Im}\{q_p\}$ at the crossover between the hydrodynamic and collisionless regime. However, while the quality of our results allows a reliable extraction of $\text{Re}\{q_p\}$, the extraction of $\text{Im}\{q_p\}$ has a larger uncertainty. On top of that, the plasmon decay due to $\text{Im}\{q_p\}$ needs to be disentangled (in the case of device 1) from decay due to the varying tip-junction distance. As a result, we do not aim to observe the transition in the damping rate measurements (Supplementary Note 2).

Motivated by the good matching between the expected hydrodynamic velocity and the experimental data at 1.84 THz, we aim to make a more complete comparison between our experiment and the theoretical model in Ref.[7]. This model needs as input the value of $\Lambda$, $\tau$, $\tau_{\text{ee}}$, $v_F$, and of the zero-th order Landau parameter[1] (which is related to the compressibility correction[9]) $F_0^s$, and allows the calculation of the plasmon wavevector at every frequency. Our best estimates of these parameters, either measured or calculated using the theory presented in Refs.[29,30] are summarized in Table S1 with details on how they are obtained given in Supplementary Notes 3,5-7. The calculated plasmon velocity according to these parameters is shown as a solid line in Fig. 3**C** for the two devices. While the trend is correct, the model predicts that only a smaller shift should be observed in our experimental range. We attribute this discrepancy mainly to an overestimate of $\tau_{ee}$ in our many-body calculation ($\tau_{ee} = 0.16$ ps for device 1 and $\tau_{ee} = 0.17$ ps for device 2) that pushes the transition to lower frequencies. To support this, we show in Fig. 3**C** the theoretical line calculated with the same parameters but with $\tau_{ee}$ reduced by a factor of two (dashed line) and five (dotted line). This shows that a better agreement is reached by assuming that the real value of $\tau_{ee}$ is reduced by around a factor of two with respect to the calculated one. This discrepancy could motivate future theoretical investigations. Further mismatch between the theoretical prediction and the experiment can be attributed to the estimate of other parameters or to mechanisms that are not captured by the simplified theoretical model in Ref.[7].

The dispersive behaviour of the dielectric environment could also introduce a frequency dependence in our experiment. However, the hBN permittivity change in our frequency range is too small to explain the effect (Supplementary Note 8) and the palladium gate electrode has a flat response in the same range (its plasma frequency being close to 10 eV)[31].

**Conclusions**

To conclude, we have experimentally demonstrated a shift in the phase velocity of acoustic graphene plasmons in a graphene sample with a very thin (2 nm) hBN spacer when the frequency is tuned from 4.25 to 1.84 THz. The same effect was not observed in a device with a thicker hBN spacer. The magnitude of the observed shift and the frequency at which the shift is happening are in qualitative agreement with the theoretical expectation for the collisionless to hydrodynamic transition.

Two main ingredients have allowed us to observe the transition. First, we have produced high mobility graphene devices in which the fastest scattering event is ee collisions. Second, we have incorporated a metallic gate electrode in very close proximity to the graphene sheet to ensure sufficient screening of the long-range Coulomb interaction, achieving record-low values of acoustic graphene plasmon velocity. Our results can stimulate further experimental investigation on the dynamical aspects of the hydrodynamic regime of electronic transport.

The ee collision rate strongly depends on temperature[10]. Performing experiments in a variable-temperature cryogenic near-field microscope[32] would permit studying the evolution of the hydrodynamic regime as a function of temperature. Moreover, the hydrodynamic regime could be studied by using THz graphene plasmon cavities coupled with a continuously-tunable THz source in the few THz range. Finally, interesting non-linear plasmonic effects are predicted to happen in the hydrodynamic regime due to the non-linearities of the Navier-Stokes equations in graphene[33,34].

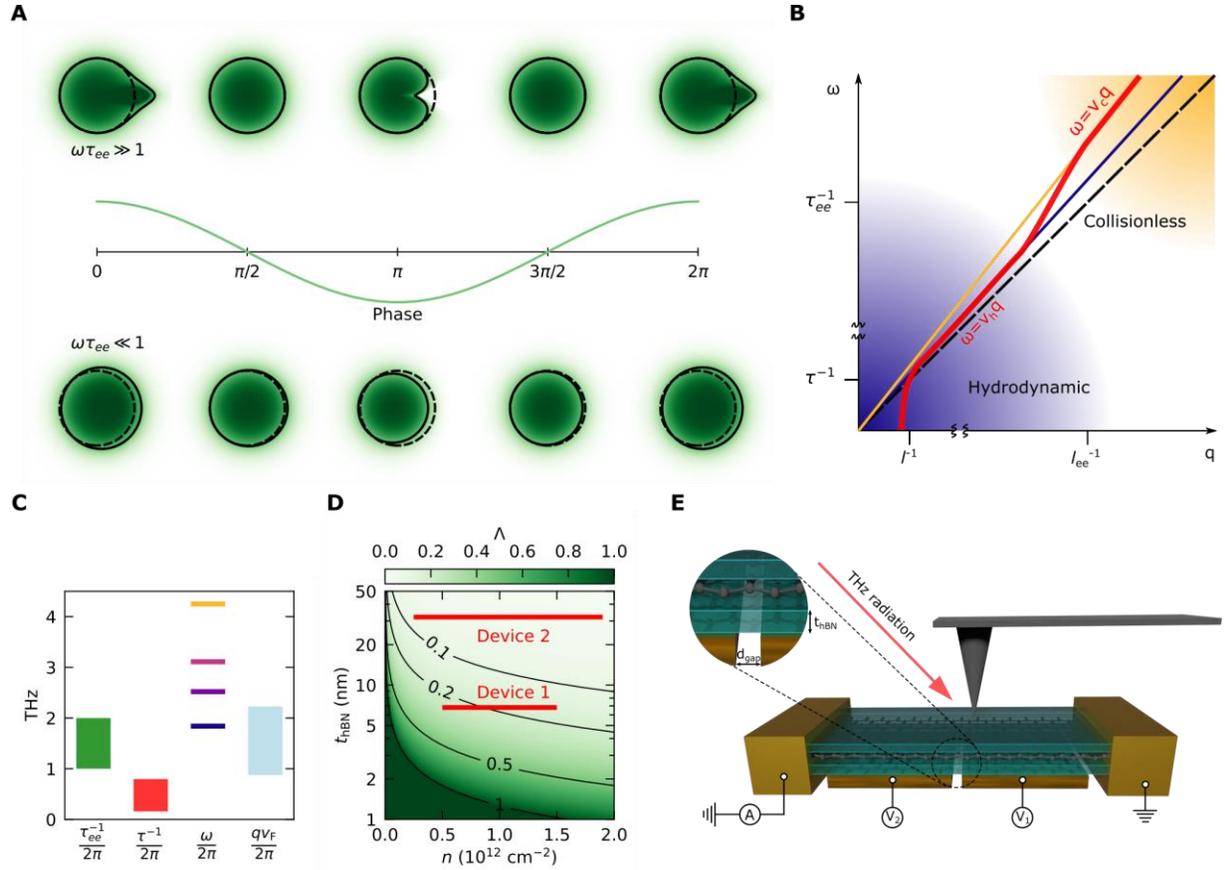

**Fig. 1. A.** Comparison of the Fermi surface deformation associated to an acoustic plasmon propagating along the positive $x$ direction between the collisionless (top row) and hydrodynamic (bottom row) regime. The green shading represents the one-particle distribution function, the black line the set of points where the distribution function is $1/2$. The dashed line marks the position where the equilibrium distribution function is $1/2$. The oscillation represents the potential energy or density perturbations. The $x$ component of the current has also (approximately) the same phase, the electric field is instead $90°$ out of phase. The phase is given by $qx - \omega t$. **B.** Dispersion of acoustic plasmons (red thick line) highlighting the change in phase velocity between the two regimes. The yellow and the blue lines represent the phase velocity in the collisionless and hydrodynamic regimes, respectively. The dashed black line corresponds to the Fermi velocity. **C.** Comparison of the frequency-scales involved in our experiment. The laser frequency $\omega$ can be tuned to be larger or smaller than the ee collision rate $\tau_{ee}^{-1}$, while the momentum relaxation rate $\tau^{-1}$ is always the slowest mechanism. $qv_F$ is always smaller than $\omega$ since the plasmon phase velocity never reaches the Fermi velocity $v_F$ for the values of the screening parameter $\Lambda$ in our experiment. $\tau_{ee}^{-1}$ has been calculated for our devices' parameters (Supplementary Note 7), while $\tau^{-1}$ has been extracted from the high-frequency data (Supplementary Note 5). **D.** Screening parameter $\Lambda$ for single layer graphene as a function of the hBN thickness $t_{hBN}$ and carrier density. Red lines represent the data ranges for our devices once the air gap is taken into account (Supplementary Note 3). **E.** Schematic view of the dual-gated device geometry and terahertz nanoscopy experiment. The inset highlights the relevant geometric dimensions.

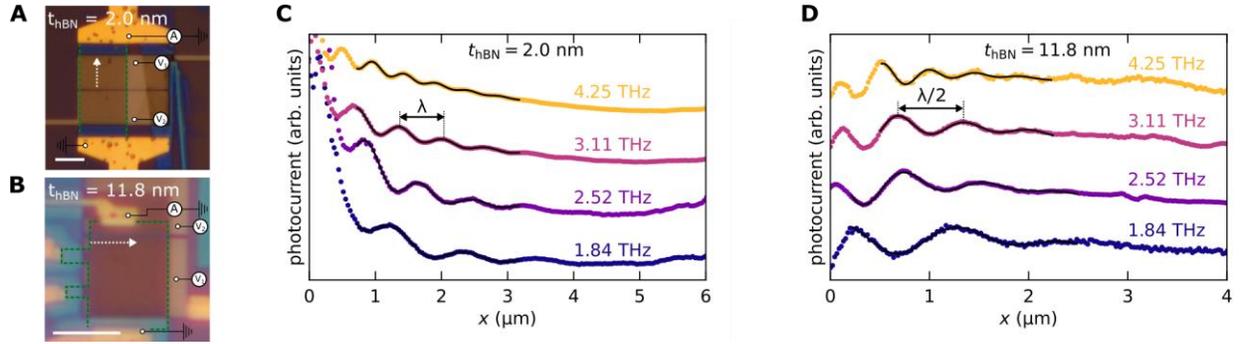

**Fig. 2. A**. Optical micrographs of device 1, indicating the electrodes used for collecting the photocurrent signal, and the gate electrodes. The white dashed line marks where datasets in **C** were acquired. The green dashed lines delimitate the area covered by graphene. The scale bar is 5 $\mu$m. **B.** Same as panel **A** for device 2, here the white dashed line marks where datasets in **D** were acquired. **C**. Near-field photocurrent signal acquired in device 1 along the line shown in **A**, at four different frequencies. The carrier density is fixed at $n \approx 10^{12}$ cm$^{-2}$. In this configuration (perpendicular to the junction) only $\lambda$-fringes appear. Data are shifted vertically for more clarity. **D**. Same as in panel **C** for device 2, along the line marked in panel **B**. In this configuration (parallel to the junction) only $\lambda/2$-fringes appear. Colour code is the same as in panel **C**.

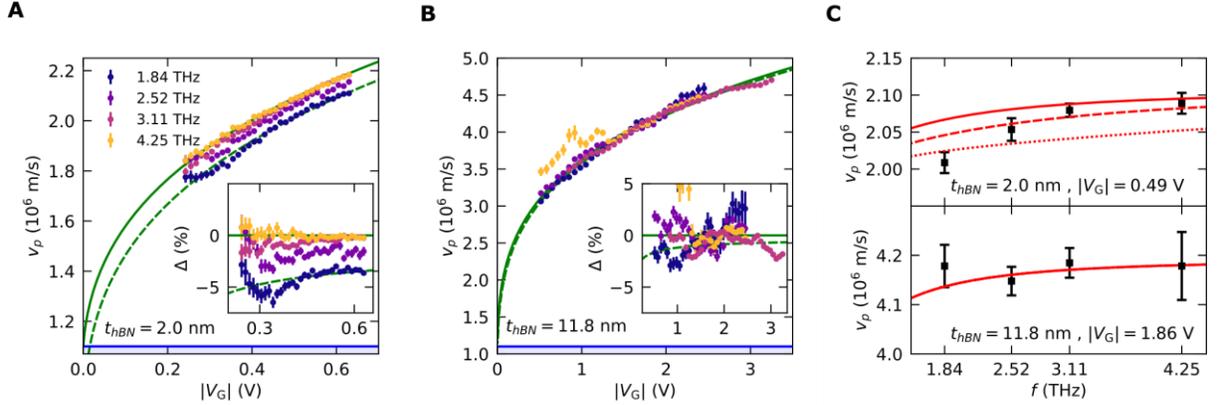

**Fig. 3. A.** Measured plasmon phase velocity as a function of carrier density for device 1, for the frequencies indicated in the legend. The green solid line is a one parameter fit $v_c(V_G)$ of the data at 4.25 THz. The green dashed line is the corresponding hydrodynamic velocity, obtained by applying Eq. 2 to $v_c(V_G)$. The inset shows the relative variation of the plasmon velocity with respect to the fit, $\Delta = v_p(V_G,\omega)/v_c(V_G) - 1$, as a function of the gate voltage. The blue shaded region indicates the area below the Fermi velocity. **B.** Same as in **A** for the control device 2. **C.** Plasmon phase velocity as a function of the frequency for device 1 (top) and device 2 (bottom), for a $V_G$ corresponding to a carrier density of $n \approx 10^{12}$ cm$^{-2}$. The data points and error bars have been obtained from fitting the dispersions in **A** and **B** to the functional form explained in the text. The solid red lines follow the expected plasmon velocity for our devices' parameters (see Table S1) using the model in Ref.[7]. The dashed and dotted red lines correspond to a value of $\tau_{ee}$ (0.16 ps for device 1 and 0.17 ps for device 2) reduced by a factor two and five respectively. The difference between the three lines is not visible for device 2.

**Acknowledgments:**

**Funding:**

Government of Spain through CEX2019-000910-S [MCIN/AEI/10.13039/501100011033] (DBR, IT, NCHH, HHS, CRM, FHLK)

Fundació Cellex (DBR, IT, NCHH, HHS, CRM, FHLK)

Fundació Mir-Puig (DBR, IT, NCHH, HHS, CRM, FHLK)

Generalitat de Catalunya through CERCA (DBR, IT, NCHH, HHS, CRM, FHLK)

Secretaria d'Universitats i Recerca del Departament d'Empresa i Coneixement de la Generalitat de Catalunya, as well as the European Social Fund (L'FSE inverteix en el teu futur)-FEDER (DBR)

European Union's Horizon 2020 programme under the Marie Skłodowska-Curie grant agreement Ref. 843830 (HHS)

European Union's Horizon 2020 research and innovation programme under the Marie Skłodowska-Curie grant agreement ref. 665884 (NCHH)

Spanish Ministry of Science, Innovation and Universities (MCIU) and State Research Agency (AEI) via the Juan de la Cierva fellowship ref. FJC2018-037098-I (IT)

ERC TOPONANOP (726001) (FHLK)

Government of Spain (PID2019-106875GB-I00) (FHLK)

Generalitat de Catalunya (AGAUR, SGR 1656) (FHLK)

European Union's Horizon 2020 under grant agreement no. 881603 (Graphene flagship Core 3) (FHLK, MP)

European Union's Horizon 2020 under grant agreement no. 820378 (Quantum flagship) (FHLK)

The University of Pisa under the "PRA - Progetti di Ricerca di Ateneo" (Institutional Research Grants) - Project No. PRA_2020-2021_92 "Quantum Computing, Technologies and Applications" (MP)

MUR - Italian Minister of University and Research under the "Research projects of relevant national interest - PRIN 2020" - Project No. 2020JLZ52N, title "Light-matter interactions and the collective behavior of quantum 2D materials (q-LIMA)" (MP)

Swiss National Science Foundation (Grant No. 172218) (CMM)

European Union's Horizon 2020 research and innovation programme under the Marie Skłodowska-Curie grant agreement No 873028 (AP)

Leverhulme Trust under the grant agreement RPG-2019-363 (AP)

JSPS KAKENHI (Grant Numbers 19H05790, 20H00354 and 21H05233) (TT, KW)

Spanish Ministry of Science and Innovation under the María de Maeztu Units of Excellence Program (CEX2020-001038-M) and the Projects RTI2018-094830-B-100 and PID2021-123949OB-I00. (RH)

**Author contributions:**

Conceptualization: IT, DBR, FHLK

Sample fabrication: DBR, NCHH

Experiment: DBR, NCHH, HHS, CMM

hBN crystals: KW, TT

Data analysis: DBR, IT, CRM

Theoretical calculations: RA, AP, IT

Supervision: IT, FHLK, MP, RH

Writing – original draft: IT, DBR, FHLK

Writing – review & editing: all authors

# Supplementary Materials for

## Transition from acoustic plasmon to electronic sound in graphene

David Barcons Ruiz, Niels C.H. Hesp, Hanan Herzig Sheinfux, Carlos Ramos Marimón, Curdin Martin Maissen, Alessandro Principi, Reza Asgari, Takashi Taniguchi, Kenji Watanabe, Marco Polini, Rainer Hillenbrand, Iacopo Torre, Frank H.L. Koppens
Correspondence to: frank.koppens@icfo.eu, iacopo.torre@icfo.eu

**Table of contents:**

**Materials and Methods**

Device fabrication

We start with the fabrication of the metallic gates. We use standard electron beam lithography (EBL) at 30 kV to define two rectangles separated by 200-300 nm gaps on a 270 nm thick polymethyl methacrylate (PMMA) layer. After developing, we perform a plasma descum at low power to remove resist leftovers. We deposit 2 nm of Ti and 15 nm of Pd, both by electron beam evaporation. Finally, to remove the spikes at the edges of the gates, we anneal the samples at 300 °C in Ar/$H_2$ for 3 hours. We check with atomic force microscopy (AFM) and choose only the gates without spikes. We find that the gap is in the order of 100-200 nm.

We mechanically exfoliate hexagonal boron nitride (hBN) and graphene flakes on SiO2/Si chips, and carefully choose the desired hBN flakes for our devices. To assemble the hBN/graphene/hBN heterostructure, we use polycarbonate (PC) stamps, and drop the heterostructure onto the pre-patterned metallic gates at 160 °C [1]. Finally, we define the edge contacts[2] and shape the graphene channel with EBL and reactive ion etching (RIE).

Prior to our nearfield measurements, we clean the surface of the samples with an AFM tip in contact mode[3], applying forces between 30 – 60 nN.

In Fig. S1 we show an AFM image of the surface of device 2 after the AFM cleaning.

Terahertz photocurrent nanoscopy measurements

As the laser source, we used two far-infrared gas lasers: FIRL-100 (Edinburgh Instruments Ltd.) and SIFIR-50 (Coherent Inc.). Both lasers output the same terahertz lines at very similar powers. As the nearfield microscope, we used a neaSNOM (neaspec GmbH). Since we perform photocurrent measurements, we removed the interferometer to maximize the power at the tip. The photocurrent signal (typically in the order of few nA) is read out through a photocurrent amplifier (DHPCA-100 from FEMTO Messtechnik GmbH), working at gains between $10^4$-$10^6$ V/A, depending on the device resistance and laser power. The amplifier output is fed to the neaSNOM lock-in input, such that the collected signal is demodulated at the harmonics of the tip frequency. We used an Au-coated AFM tip with 250 nm radius at the apex and 3 N/m force constant, model LRCH250 (Team Nanotec GmbH). The photocurrent signal is demodulated at either the first or second harmonic of the tip frequency (~ 75 KHz), and the first harmonic of the mechanical phase is substracted[4,5]. The typical tapping amplitude is 80-120 nm.

First, we locate a clean line perpendicular to the junction for device 1 and parallel to it, but close enough to maximize the signal, for device 2. Whether we want to measure on the electron or hole doping regimes, we choose a different gate voltage for the other gate electrode to maximize the photocurrent. We scan along the same line, for a range of gate voltages. In Fig. S2 we display the raw measurements acquired for device 1 (left) and device 2 (right), where in the horizontal axis we scan the tip and in the vertical axis we step the gate voltage.

We check for position and carrier density drifts between scans that may alter the data. We always scan across the pn junction (device 1) or across the graphene edge (device 2). This, together with comparing forward and backward traces (which are recorded sequentially), allows us to discard sample drifts that could lead to an apparent change in the fringe spacing. To check for carrier density drifts, i.e. a drift of the charge neutrality point $V_{CNP}$, we verify that the gate voltage at which the photocurrent signal changes its sign remains the same. Moreover, we do not expect this to happen in samples with a local gate. This samples are much less affected by drift and hysteresis than encapsulated samples directly on top of SiO2 due to the lack of dielectric-dielectric interfaces that may trap charges for long times.

**Supplementary Text**

1. Plasmon launching: tip or junction

In our experiment, we can distinguish two types of plasmon launching in our devices (Fig. S3): tip-launched plasmons, which appear as λ/2-fringes, and junction-launched plasmons, appearing as λ-fringes in the experiment.
- In the case of tip-launched plasmons (λ/2-fringes), the tip provides the momentum mismatch to launch the plasmon in all directions (circular wave). Once it reaches the lithographically defined graphene edge, it is reflected and travels back to the tip.
- In the case of junction-launched plasmons (λ-fringes), the launching happens at the sharp edge of the metallic gate. A sharp metallic edge can compensate the momentum mismatch between the incoming light and the plasmon-polariton, similar to what the SNOM tip does in most experiments. Due to the large length of the junction compared to the plasmon wavelength, we consider the plasmon as a plane wave propagating perpendicular to the junction.

2. Plasmon fringe fitting

As mentioned in the main text and in the Supplementary Note 1, we have two different plasmon launching mechanisms for device 1 and 2. Therefore, the fringe fitting strategy of the plasmon oscillations slightly differs.
- In device 1 ($t_{hBN} = 2.0$ nm), we observe only junction-launched plasmons (λ-fringes). The tip, which is scanned perpendicular to the junction, acts as a rectifier, thus converting the plasmon into heat[4]. The heat is dissipated through the hBN[6] and in the graphene layer, rising the electronic temperature at the junction side where we rectify the plasmon. Since the Seebeck coefficient is tuned to different values at both sides of the junction, the temperature difference results in a thermo-voltage generation. The voltage difference generates a current flow which is the signal we measure. In summary, we need to consider λ-fringes which propagate over a characteristic distance $l_p$, being carried by a signal that decays nearly exponentially away from the junction, with a characteristic decay length $l_T$ (cooling length). The latter should also be included as an extra decay channel for the oscillations as we scan away from the junction. We define $k_p^\dagger$ as the complex wavevector of the measured fringes. This is related to the plasmon wavevector by $\text{Re}\{k_p^\dagger\} = \text{Re}\{k_p\} = 2\pi/\lambda_p$ and $\text{Im}\{k_p^\dagger\} \approx 1/l_p + 1/l_T$, where the second term accounts for the loss of heat related to the cooling length. The fitting function used read as

$$s(x) = Ae^{ik_p^\dagger x} + Be^{-x/l_T} + C \qquad (S1)$$

where $A, B, C$ are amplitude coefficients.
- In device 2 ($t_{hBN} = 11.8$ nm), we did not observe λ-fringes (with the exception of small voltage data at 4.25 THz), which might be due to its thicker bottom hBN flake. Instead, we do observe λ/2-fringes, i.e. tip-launched plasmons. The tip, which is scanned parallel to the junction, acts as a rectifier the same way it does for device 1, and we do not expect any decay of the signal due to the cooling length. Next to the graphene edges, the heat diffusion is altered due to the discontinuity. Thus, the collected photocurrent will vary close to the edge, so we account for the background signal $s_{bg}(x)$ with a fourth order polynomial. The signal is fitted with the following function:

$$s(x) = s_{bg}(x) + (A + iB)e^{i2k_p x} \qquad (S2a)$$

where $\text{Im}\{k_p\} = 2\pi/l_p$, and $A$ and $B$ amplitude coefficients. At 4.25 THz we do also observe at small voltages λ-fringes. In order to obtain an accurate value of the decay for the plasmon and carrier lifetime analysis (Supplementary Note 5), we need to account for both oscillations and for the geometrical decay (see Ref. [7] for more details). We use the following function, where $s_{bg}(x)$ is a third order polynomial:

$$s(x) = s_{bg}(x) + \frac{A}{\sqrt{x}}\cos(2\text{Re}\{k_p\}x + \varphi)e^{-2\text{Im}\{k_p\}x} + \frac{B}{x}\cos(\text{Re}\{k_p\}x + \theta)e^{-\text{Im}\{k_p\}x} \qquad (S2b)$$

where $A$ and $B$ amplitude coefficients, and $\varphi$ and $\theta$ angles to account for the global and relative phase of the two oscillations.

In both cases the fringe fitting was performed excluding the first oscillation that may be affected by edge effects. For device 1, the imaginary part of the plasmon wavevector is affected by a larger error due to the subtraction of two contributions. In Fig. S4 we show $\text{Re}\{k_p\}$, and $\text{Im}\{k_p\}$ (for device 1 we plot $\text{Im}\{k_p^\dagger\}$), extracted from the fits, as a function of frequency and gate voltage $V_1$.

3. Capacitance and carrier density determination

To convert the gate voltage $V_1$ to carrier density according to $n = -eC_{tot}(V_1 - V_{1,CNP})$, we include in the total capacitance the capacitance of an hBN layer with a known thickness $t_{hBN}$ from AFM, an air gap $t_{air}$ between the metallic gate and the hBN due to possible contamination or surface roughness, and the quantum capacitance $C_q(n)$ of graphene. This gives

$$\frac{1}{C_{tot}} = \frac{1}{C} + \frac{1}{C_q} = \frac{1}{\frac{\epsilon_0 \epsilon_{hBN}}{t_{hBN}}} + \frac{1}{\frac{\epsilon_0}{t_{air}}} + \frac{1}{\frac{e^2}{2}DoS(n)}, \qquad (S3)$$

where $\epsilon_0$ is the vacuum permittivity, $\epsilon_{hBN}$ is the out of plane dielectric constant of hBN ($\epsilon_{zz,hBN} = 3.56$) and $DoS(n) = 2\sqrt{|n|}/(\sqrt{\pi}\hbar v_F)$ is the density-dependent density of states in graphene. With this relation and the parameters $\tau, \tau_{ee}, F_0, v_F$ estimated or calculated (See following Supplementary Notes) the relation between voltage and plasmon wavelength given in Ref.[8] reduces to a model with one fit parameter, namely the thickness of the air layer.

In order to find this value, we fit the plasmon dispersion at 4.25 THz to the experimental data. We find $t_{air}^{device\,1} \approx 1.46$ nm and $t_{air}^{device\,2} \approx 6.24$ nm. This is in good agreement with the AFM imaging of the device, where we found $t_{air}^{device\,1} = (1.0 \pm 0.5)$ nm and $t_{air}^{device\,2} = (5.0 \pm 1.0)$ nm. The latter values present a large uncertainty due to the indirectness of the measurement. In fact, we first measured the thickness of the flakes before patterning the contacts and defining the channel. After the measurements, we measure the step height from the metallic gate to the top of the heterostructure (Fig. S5). Therefore, we could extract $t_{air}$ from the measured step, $\Delta z = t_{air} + t_{hBN} + t_G + t_{top-hBN}$ and from the independent AFM measurements for hBN flake thicknesses $t_{top-hBN}^{device\,1} = 10.8$ nm and $t_{top-hBN}^{device\,2} = 15.0$ nm (we assume $t_G = 0.3$ nm).

## 4. Plasmon phase velocity in the collisionless and hydrodynamic regime

The plasmon phase velocity in the collisionless and hydrodynamic regime are given in Eqs. (10-12) of Ref. [8]. These two formulas drastically simplify if the Landau parameter of order one $F_1^s$ is neglected. This is a good approximation in single-layer graphene since $F_1^s$ is not constrained by Galileian invariance (allowing a many-body renormalization of the Drude weight) and the expected values of $F_1^s$ are small[8]. After making this approximation, expressing the many-body renormalizations in terms of Landau parameters, and adapting to the notations used in this work the two equations read

$$v_c = S_c = v_F \frac{1 + \Lambda^{-1} + F_0^s}{\sqrt{1 + 2\Lambda^{-1} + 2F_0^s}}, \qquad (S4)$$

$$v_h = S_h = v_F \sqrt{\frac{1 + \Lambda^{-1} + F_0^s}{2}}. \qquad (S5)$$

It must be noted that here $v_F$ is the observed, i.e. renormalized Fermi velocity, corresponding to $v_F^*$ in Ref. [8], moreover $\Lambda$ is defined using the renormalized density of states and corresponds to the quantity $(v_F^*/v_F)\Lambda$ in the notation used in Ref. [8].

In our work we define the plasmon velocity as the phase velocity of the mode, i.e. $v_p(\omega) = \omega/Re[q_p(\omega)]$. Reference[8] defines instead the plasmon velocity as $S_\omega = \omega/\sqrt{Re[q_p^2(\omega)]}$, the two definitions converging for small dissipation ($Im[q_p(\omega)] \ll Re[q_p(\omega)]$). In both the collisionless and hydrodynamic limit $\omega \gg \tau^{-1}$ and $v_p$ coincides with $S$. We also checked that in all our datasets the difference between $v_p$ and $S$ is negligible thanks to the high quality of our samples.

Solving (S4) for the quantity $1 + \Lambda^{-1} + F_0^s$ and substituting back into (S5) yields the relation in Eq. (2) of the main text.

## 5. Plasmon and carriers lifetimes

For the highest frequency (4.25 THz) the imaginary part of the plasmon wavevector can be extracted. This allows to calculate the plasmon damping rate according to $\Gamma_\omega = (\omega Im[q_p^2(\omega)])/(2Re[q_p^2(\omega)])$ as in Ref. [8]. The results are shown in Fig. S6, and yield plasmon lifetimes (inverse of the damping rate) of the order of 0.5-1 ps.

We note that this definition of the plasmon damping rate differs slightly from the one used in most of the plasmonic literature $\Gamma(\omega) = (\omega Im[q_p(\omega)])/(Re[q_p(\omega)])$. Again, the two definitions converge for $Im[q_p(\omega)] \ll Re[q_p(\omega)]$ and the difference is irrelevant in all our datasets.

The plasmon damping rates in the collisionless and hydrodynamic regime are given in Eq. (11-13) of Ref. [8]. After making the same approximation and change of notations as discussed in the previous section we get

$$\Gamma_c = \tau^{-1} \frac{1 + \Lambda^{-1} + F_0^s}{1 + 2\Lambda^{-1} + 2F_0^s} + \tau_{ee}^{-1} \frac{1 + \Lambda^{-1} + F_0^s}{(1 + 2\Lambda^{-1} + 2F_0^s)^2}, \qquad (S6)$$

$$\Gamma_h = \frac{\tau^{-1}}{2} + \frac{\omega^2}{4(\tau^{-1} + \tau_{ee}^{-1})(1 + \Lambda^{-1} + F_0^s)}. \qquad (S7)$$

Where we remember that $\tau$ is the momentum relaxation scattering time i.e. the time that is extracted from mobility measurements. In the simple with low screening ($\Lambda \ll 1$) both formulas reduce to

$$\Gamma = \frac{\tau^{-1}}{2}. \qquad (S8)$$

To extract the momentum relaxation time $\tau$ we use the following procedure. We assume that at 4.25 THz the collisionless limit is fully developed and we can use formulas (S4-S6). At each voltage point we calculate the quantity $1 + \Lambda^{-1} + F_0^s$ from (S4) using the measured value of the plasmon velocity and a fixed value of $v_F = v_F(n = 10^{12} \text{ cm}^{-2}) = 1.1 \cdot 10^6 \, m/s$. We then substitute in (S6), and using the measured value of $\Gamma$ (Fig. S7) and the fixed value of $\tau_{ee} = \tau_{ee}(n = 10^{12} \text{ cm}^{-2}) = 0.16$ ps we can find $\tau$. The results are shown in Fig. S7 together with the estimation resulting from the simple application of (S8) that offers a lower bound for $\tau$. Both values are in the range of expected mobility scattering times for this type of structures and choosing one or the other does not alter our conclusions.

6. Many body renormalization corrections

The many-body renormalization of the velocity can be calculated in graphene from the real part of the electronic self-energy according to the theory developed in Ref.[9]. In Fig. S8 we show the numerical results obtained taking into account the electrostatic screening in the precise geometry of our heterostructures, including the air gaps discussed in Supplementary Note 3. The density dependence is weak and almost no difference is observed between the two devices.
A calculation of the random-phase-approximation (RPA) ground-state energy of the graphene electron liquid allows extracting the renormalized compressibility as described in Ref.[9]. This, together with the renormalization of the Fermi velocity allows to extract the spin-symmetric Landau parameter of order zero $F_0^s$ that enters in the expressions for the velocity of the plasmon mode. In Fig. S9 we show the results of the numerical calculation for our device geometries.

7. Electron-electron scattering time

The electron-electron scattering time $\tau_{ee}$ can be extracted from a calculation of the imaginary part of the $G_0W$ self-energy of electrons in graphene[10]. In Fig. S10 we report the results of numerical calculations for our relevant electronic density range and room temperature $T = 295$ K, taking into account the realistic dielectric screening in the heterostructures, including the air gaps discussed in Supplementary Note 3.

8. hBN permittivity frequency dependence

To discard that the observed effects could be originated by the frequency dependent hBN dielectric functions, we plot in Fig. S11 its in-plane $\varepsilon_\parallel$ and out-of-plane $\varepsilon_\perp$ permittivities as a function of frequency following the model in Ref. [11].
In the frequency range of our experiment, i.e. from 1.84 THz to 4.25 THz, the relative change of the real part of the hBN permittivities is $\Delta\varepsilon_\parallel/\varepsilon_\parallel = 0.24$ % and $\Delta\varepsilon_\perp/\varepsilon_\perp = 0.52$ %.

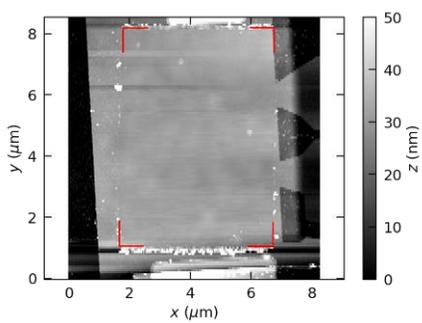

**Fig. S1.** AFM topography image of device 2 after AFM cleaning in contact mode the region inside the red marks.

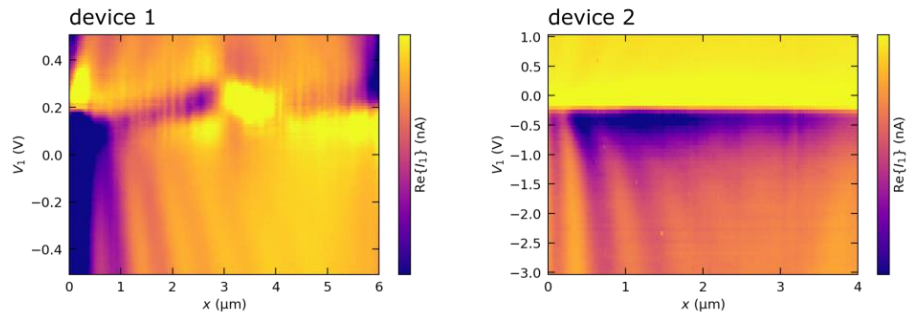

**Fig. S2.** Photocurrent signal $I_1$ as a function of gate voltage $V_1$ and position $x$, for device 1 (left panel) and device 2 (right panel). The laser frequency was 2.52 THz.

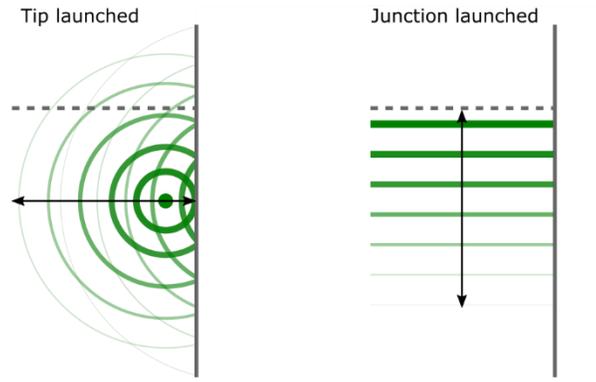

**Fig. S3.** Plasmon launching mechanisms. The green lines and circles indicate the plasmon wavefront. The black arrows indicate the scanning direction in the experiment. For tip-launched, it results in λ/2-fringes, while for junction-launched in λ-fringes. The solid grey line corresponds to the graphene edge and the dashed grey line to the location of the junction.

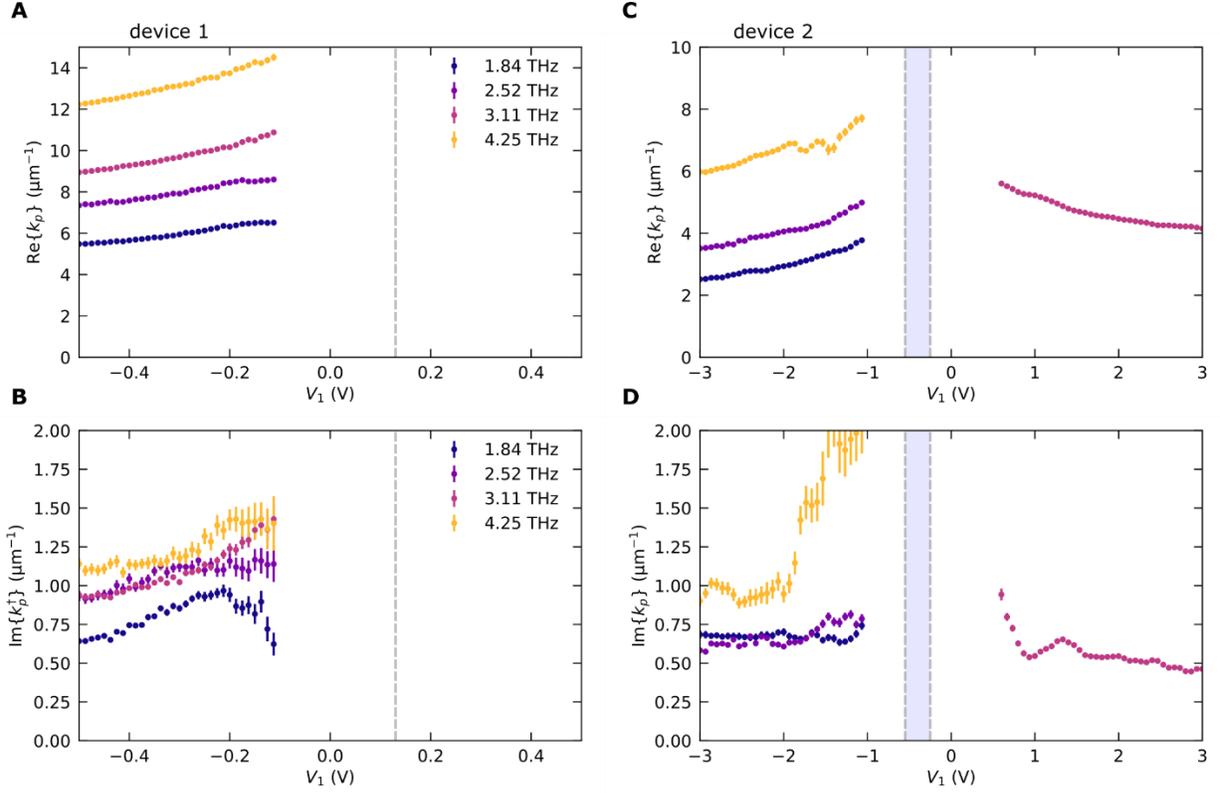

**Fig. S4. A,B.** Extracted Re{$k_p$} and Im{$k_p^\dagger$} as a function of gate voltage $V_1$, for device 1. **C,D.** Extracted Re{$k_p$} and Im{$k_p$} as a function of gate voltage $V_1$, for device 2. Vertical dashed lines indicate the position of charge neutrality point during the measurements, extracted from the two-terminal resistance. For device 2, there was a small shift during the measurements (considered in the analysis), indicated by the blue-shaded region.

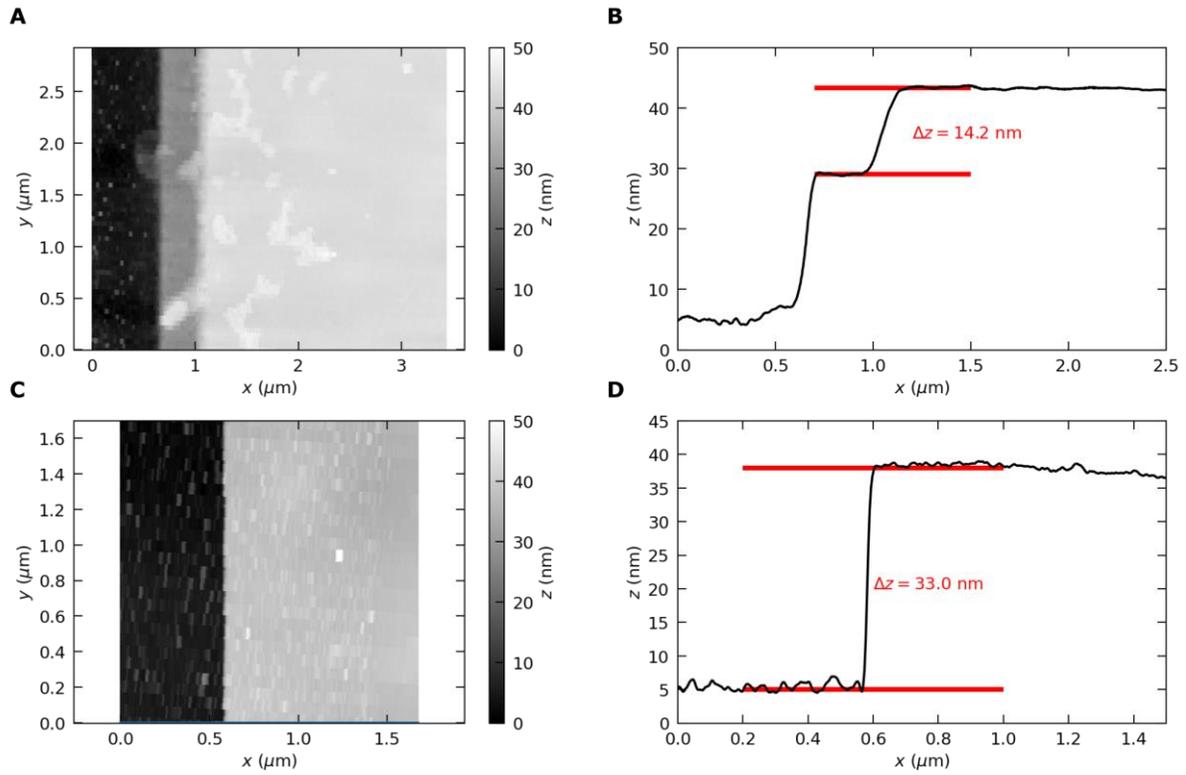

**Fig. S5**. **A.** AFM image of device 1 step from the metallic gate to the top of the hBN/graphene/hBN heterostructure. **B.** Step height profile extracted by averaging the AFM image in a. Δ$z$ indicates the total step height including the air gap discussed in the text. **C,D.** Same as **A,B** but for device 2, respectively.

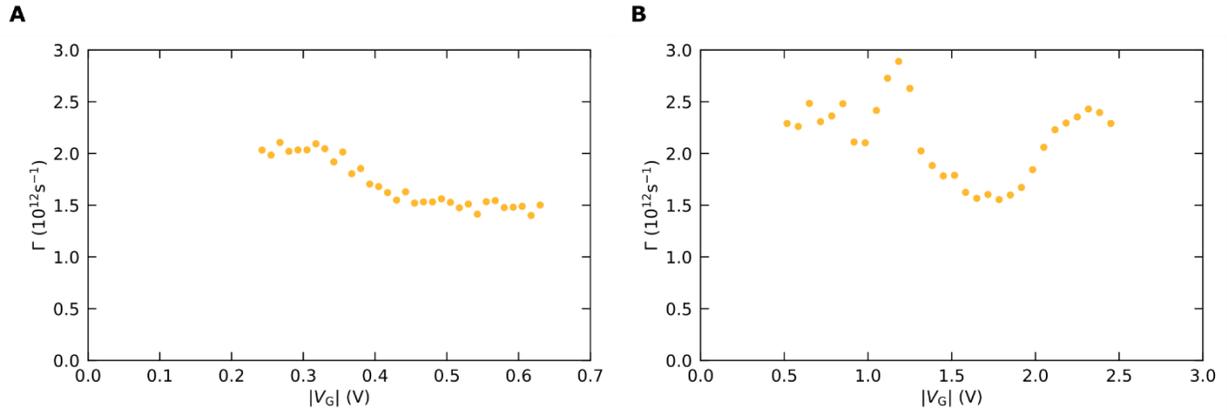

**Fig. S6. A.** Plasmon damping rate Γ extracted from data at 4.25 THz in device 1. **B.** Same as in **A** for device 2.

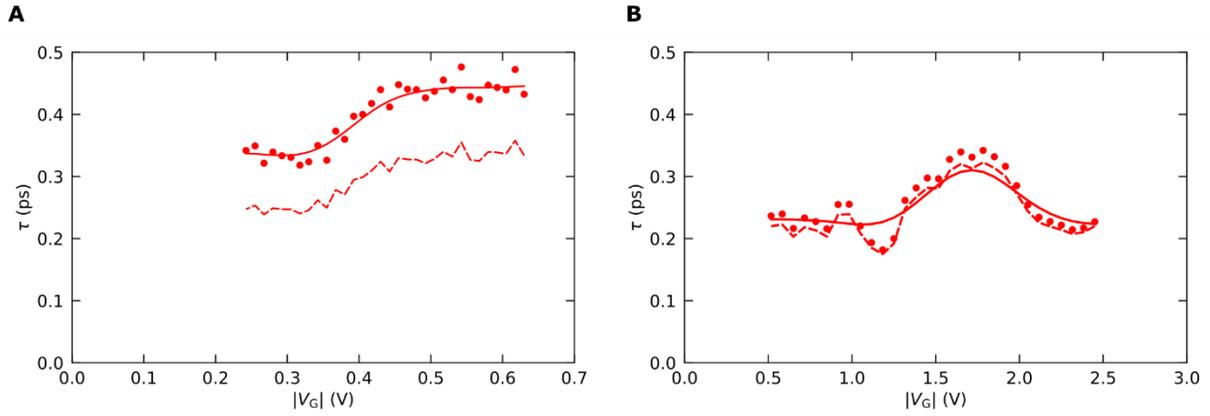

**Fig. S7. A.** Scattering time for momentum-relaxing collisions as a function of the gate voltage for device 1. Red dots are the result of the application of the procedure explained above, the dashed line results from the application of (S8), while the solid line is a smoothening of the experimental points. **B.** Same as in **A** for device 2.

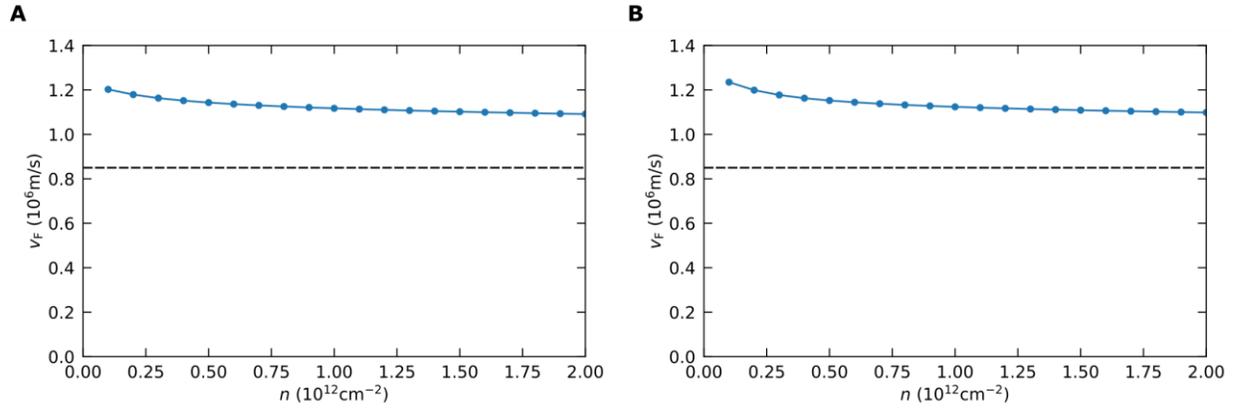

**Fig. S8. A.** Calculated many-body renormalized Fermi velocity in single-layer graphene, as a function of electronic density, taking into account the realistic dielectric environment of the heterostructure including the air gap for device 1. The black dashed line marks the bare value of the Fermi velocity $0.85 \cdot 10^6$ m/s. **B.** Same as in **A** for device 2.

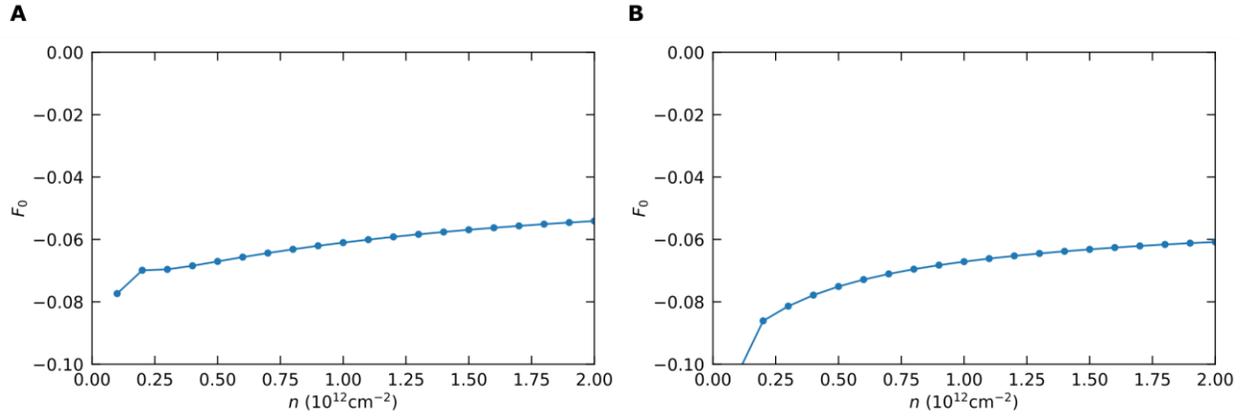

**Fig. S9. A.** Calculated spin-symmetric Landau-Fermi liquid parameter $F_0$ for single-layer graphene, as a function of electronic density, taking into account the realistic dielectric environment of the heterostructure including the air gap. **B.** Same as in **A** for device 2.

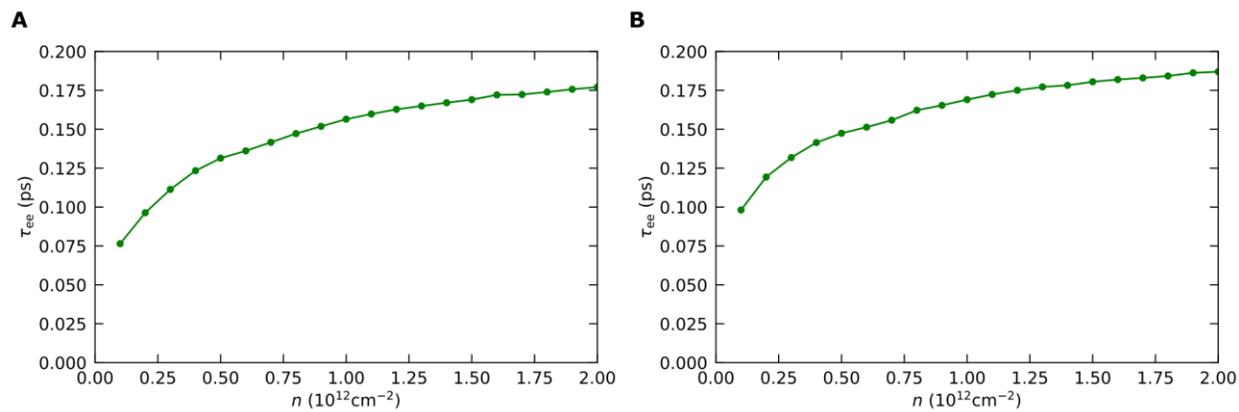

**Fig. S10. A.** Calculated electron-electron scattering time for single-layer graphene at room temperature, as a function of electronic density, taking into account the realistic dielectric environment of the heterostructure including the air gap. **B.** Same as in **A** for device 2.

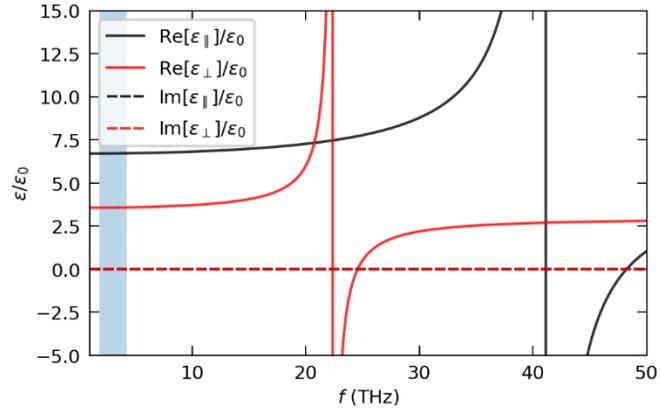

**Fig. S11.** hBN permittivity as a function of frequency, based in the model in Ref.[11]. All permittivities are normalized to the vacuum permittivity. Blue shaded region indicates the frequency range in our experiment.

**Table S1.** Summary of parameter used in Fig. 3C in the main text.

| Parameter | Device 1 | Device 2 | Source |
|---|---|---|---|
| Air gap thickness $t_{air}$ | 1.46 nm | 6.24 nm | Fitted (See Supp. Note 3) |
| Capacitance $C$ | 4.37 mF/m² | 0.93 mF/m² | Extracted from $t_{air}$ (See Supp Note 3) |
| Gate voltage $V_G$ | 0.49 V | 1.86 V | Measured |
| Density $n$ | $10^{12}$ cm$^{-2}$ | $10^{12}$ cm$^{-2}$ | Inferred from the two values above |
| Screening parameter $\Lambda$ | 0.178 | 0.038 | Calculated with Eq. (1) with the renormalized Fermi Velocity |
| Fermi Velocity $v_F$ | 1.1 10⁶ m/s | 1.1 10⁶ m/s | Calculated (See Supp. Note 6) |
| Zeroth-order Landau parameter $F_0^s$ | −0.061 | −0.067 | Calculated (See Supp. Note 6) |
| Momentum relaxation time $\tau$ | 0.44 ps | 0.30 ps | Measured (See Supp. Note 5) |
| Electron-electron scattering time $\tau_{ee}$ | 0.16 ps | 0.17 ps | Calculated (See Supp. Note 7) |